\documentclass[12pt,aps,prb,preprint]{revtex4}

\usepackage{amsmath}
\usepackage{graphicx}

\begin{document}

\title{An accurate formula for the period of a simple pendulum
oscillating beyond the small angle regime}

\author{F. M. S. Lima}
\email{fabio@fis.unb.br}
\affiliation{Instituto de F\'{\i}sica, Universidade de Bras\'{\i}lia, P. O.
Box 04455, 70919-970, Bras\'{\i}lia-DF, Brazil}

\author{P. Arun}
\affiliation{Department of Physics and Electronics, SGTB Khalsa College,
University of Nova Delhi, Delhi 110 007, India}


\begin{abstract}
A simple approximate formula is derived for the dependence of the
period of a simple pendulum on the amplitude. The approximate is
more accurate than other simple formulas. Good agreement with
experimental data is verified.
\end{abstract}

\maketitle

\section{Introduction}
The periodic motion exhibited by a simple pendulum is harmonic only
for small angle oscillations.\cite{Serway} Beyond this limit, the
equation of motion is nonlinear. Although an integral formula exists
for the period of the nonlinear pendulum, it is usually not
discussed in introductory physics classes because it is not possible
to evaluate the integral exactly.\cite{Marion} For this reason
almost all introductory physics textbooks and lab manuals discuss
only small angle oscillations for which the approximation
$\sin\theta \approx \theta$ is valid. The linearized equation has a
simple exact solution, whose derivation can be easily understood by
first-year students.\cite{Serway} This linearization has bothered us
since our own undergraduate days because the amplitude needs to be
less than $7^{\circ}$ if an error less than 0.1\% (the typical
experimental error obtained with a stopwatch) is desired.
Measurements in undergraduate labs rarely have such small
amplitudes,\cite{exception} and interested students sometimes ask
for a relation that can describe the increase of the period observed
for large amplitudes.\cite{Fulcher}

The restriction to small angle oscillations hinders the
understanding of real-world behavior because the pendulum
isochronism observed in the small angle regime vanishes for
increasing amplitudes. This restriction is also unnecessary because
millisecond precision in measurements of the period is easily
obtained with current technology.\cite{Arun,Araki,Moreland,Kidd} For
instance, an experimental error of the order of 0.1\% or less is
typically obtained with a one meter-long pendulum, and thus accurate
experimental studies of the dependence of the period on amplitude
are possible even in introductory physics labs.\cite{Moreland,Kidd}

In this paper we derive a simple and accurate formula for the period
of a pendulum oscillating beyond the small angle regime. The
deviation from the exact results are of the same order of the
experimental error.

\section{Approximation}
An ideal simple pendulum consists of a particle of mass $m$ suspended by a
massless rigid rod of length $L$ that is fixed at the upper end such that
the particle moves in a vertical circle. This simple mechanical system
oscillates with a symmetric restoring force (in the absence of dissipative
forces) due to gravity, as illustrated in Fig.~1. Its equation of motion is given by\cite{Marion}
\begin{equation}
\frac{d^2\theta}{dt^2} + \frac{g}{L} \sin \theta = 0 ,
\end{equation}
where $\theta$ is the angular displacement in radians ($\theta = 0$
at the equilibrium position) and $g$ is the local acceleration of
gravity. For any given initial condition, the exact solution can
only be obtained numerically (with arbitrary accuracy). For small
angle oscillations, the approximation $\sin\theta \approx \theta$ is
valid and Eq.~(1) becomes a linear differential equation analogous
to the one for the simple harmonic oscillator. In this regime the
pendulum oscillates with a period $T_{0} = 2\pi \sqrt{{L/g}}$, a
well-known textbook relation.\cite{Serway} This relation
underestimates the exact period for any amplitude, but the
difference is almost imperceptible for small angles. For larger
angles $T_{0}$ becomes more and more inaccurate for describing the
exact period and Eq.~(1) can be used to obtain a numerical solution.

Alternatively, an integral expression
for the exact pendulum period may be derived from energy considerations,
without a detailed discussion of differential equations. If we take the
zero of potential energy at the lowest point of the trajectory (see Fig.~1)
and choose for simplicity the initial conditions as
$\theta(0)=+\theta_{0}$ and $d\theta/dt(0) = 0$, we have\cite{Marion}
\begin{equation}
m g L (1-\cos\theta_{0}) = \frac{1}{2} m L^2 (\frac{d\theta}{dt})^2 + m g L
(1-\cos\theta) .
\end{equation}
The solution for $d\theta/dt$ is
\begin{equation}
\frac{d\theta}{dt} = \pm\sqrt{\frac{2g}{L}(\cos\theta-\cos\theta_{0})} ,
\end{equation}
where the $+$ ($-$) sign is for counter-clockwise (clockwise)
motion. Now, integrating $d\theta/dt$ from $\theta_{0}$ to 0 (thus
choosing the $-$ sign in Eq.~(3)), which corresponds to a time equal
to one quarter of the exact period $T$, we have
\begin{equation}
\label{integ}
T=2\sqrt{2} \sqrt{\frac{L}{g}} \!\int_{0}^{\theta_{0}}
\frac{1}{\sqrt{\cos\theta-\cos\theta_{0}}}\, d\theta.
\end{equation}
The definite integral in Eq.~\eqref{integ} cannot be expressed in
terms of elementary functions.\cite{Marion,Abramow} Note that the
numerical evaluation of the period using Eq.~\eqref{integ} is not
straightforward because the integrand has a vertical asymptote at
$\theta = \theta_{0}$, which makes the integral
improper.\cite{Schery} This difficulty can be circumvented by
substituting $\cos\theta$ by $1-2\sin^2(\theta/2)$ and making a
change of variables given implicitly by $\sin\varphi =
\sin(\theta/2)/\sin(\theta_{0}/2)$. In this way, Eq.~(4) becomes
\begin{equation}
T=4\sqrt{\frac{L}{g}} \!\int_{0}^{\pi/2}
\frac{1}{\sqrt{1-k^2\sin^2\varphi}}\, d\varphi ,
\label{T}
\end{equation}
where $k\equiv\sin(\theta_{0}/2)$. The definite integral is $K(k)$, the
complete elliptic integral of the first kind, which is not improper because
$k<1$ for $|\theta_{0}|<\pi$.

It is not difficult to numerically evaluate $T$ for a given
amplitude. The relative error made in approximating $T$ by $T_{0}$,
where $T=2T_{0}/\pi K(k)$, is\cite{obs1}
\begin{equation}
\frac{T_{0}-T}{T} = \frac{\pi}{2 K(k)}-1 .
\end{equation}

Our proposed approximation for the pendulum period is based on the
observation that
$f(\varphi,k)\equiv\sqrt{1-k^2\sin^2\varphi}$ is a smooth
function of $\varphi$, whose concavity changes from downward to upward at a point near the middle of
the interval of integration, that is, $0\leq\varphi\leq{\! \pi/2}$. As
shown in Fig.~2, this change occurs for any
$\theta_{0}$ between $0$ and ${\! \pi/2}$.\cite{obs2} We use the points
$(0,1)$ and $({\! \pi/2},a)$ for a linear interpolation, where
$a \equiv f(\varphi = \pi/2,k=
\sin\theta_0/2)=\sqrt{1-(\sin\theta_0/2)^2}=\cos \theta_{0}/2$, and
approximate $f(\varphi,k)$ by
\begin{equation}
f(\varphi,\theta_{0}) \approx 1-\frac{2}{\pi}(1-a) \varphi.
\label{r}
\end{equation}
We substitute Eq.~(\ref{r}) in the denominator of the integrand in Eq.~(\ref{T}) and find
\begin{equation}
\label{thisexp} K(k) \approx \!\int_{0}^{\pi/2}
\frac{1}{1-(2/\pi)(1-a)\varphi} d\varphi =-\frac{\pi}{2} \frac{\ln
a}{1-a}.
\end{equation}
Finally, by substituting Eq.~\eqref{thisexp} in Eq.~(5), we found
\begin{equation}
T_{\log} = - 2\pi\sqrt{\frac{L}{g}} \frac{\ln a}{1-a} = -T_0
\frac{\ln a}{1-a}. \label{eq:log}
\end{equation}

Note that $\ln a < 0$ and hence $T_{\log} > 0$ for $|\theta_{0}|<\pi$.
The relative error in the logarithmic formula in Eq.~(9) is given
by\cite{obs3}
\begin{equation}
\frac{T_{\log}-T}{T} = \frac{\pi}{2 K(k)} \frac{(-\ln a)}{1-a}-1.
\end{equation}

\section{Comparison with other approximations\label{sec:compare}}
We compare the accuracy of our approximation for the pendulum period
in Eq.~(\ref{eq:log}) in representing the exact period to that of
other known approximations for amplitudes less than or equal to
$\pi/2$.\cite{obs2} The relative errors found by approximating the
exact period by $T_{0}$ and $T_{\log}$, as well as by other
formulas, are depicted in Fig.~3, where it is easily seen that all
approximations present the same general behavior: For small
amplitudes their corresponding error curves tend to zero and for
larger amplitudes the curves go up monotonically, reflecting the
increase of the relative error with the amplitude obtained with all
known approximation formulas. However, note that the rate at which
the error increases is different for each curve, the smaller rate
being the desired one for a good approximation. In this sense, the
small angle approximation $T \approx T_{0}$ exhibits the worst
behavior because its error becomes greater than 0.1\%\,(0.5\%) above
an amplitude as low as $7^{\circ}$ ($16^{\circ}$).

The second-order approximation found by Bernoulli in 1749 from a
perturbative analysis of Eq.~(5), perhaps the most famous formula
for the large angle period, is \cite{Bernoulli}
\begin{equation}
T_{2}=T_{0}\Big(1+\frac{{\theta_{0}}^{2}}{16}\Big) .
\end{equation}
As seen from the short-dashed line in Fig.~3, it leads to an error
that increases rapidly, overcoming the 0.1\%\,(0.5\%) level for
amplitudes above $41^{\circ}$\,($60^{\circ}$). The addition of more
terms improves the accuracy of $T_{2}$.\cite{Parwani}

More recently, other approximation formulas have been proposed.
Among them, the Kidd-Fogg formula has attracted much interest due to
its simplicity.\cite{Kidd} It is given by
\begin{equation}
T_{KF}=T_{0} \frac{1}{\sqrt{\cos(\theta_0/2})} .
\end{equation}
The dash-dotted line in Fig.~3 represents the error for $T_{KF}$.
The error is greater than 0.1\% only for amplitudes $\theta_0 \geq
57^{\circ}$ and reaches 0.8\% for $\theta_0 = 90^{\circ}$. Thus, it
is not accurate enough for interpreting the experimental data for
very large-angle amplitudes, contrary to the claim of
Millet.\cite{Millet}

Another formula for the period arises when an interpolation-like
linearization is made directly in Eq.~(1).\cite{Molina} The resulting expression is
\begin{equation} T_{M}=T_{0} \Big(\frac{\sin
\theta_{0}}{\theta_{0}}\Big)^{\!-3/8} ,
\end{equation} which has an error greater than 0.1\% only for
$
\theta_0 \geq 69^{\circ}$ (see the thin solid curve in Fig.~3). However, the error
 reaches $\sim 0.4\%$ for $\theta_0 = 90^{\circ}$, which is
four times the typical experimental error (0.1\%).

The error using Eq.~\eqref{eq:log} for the period (see the thick
solid line in Fig.~3) remains below all other error curves for any
$\theta_0$ and is greater than 0.1\% only for amplitudes greater
than $74^{\circ}$. Moreover, it increases slowly, reaching only
0.2\% for $\theta_0=86^{\circ}$. Therefore, our logarithmic formula
is the better approximation for the exact pendulum period in the
sense it yields the smaller relative error and also the smaller rate
for the relative error increasing in the range of amplitudes studied
here.

\section{Experiment and results}
Reliable data for large-angle pendulum periods were obtained by
Fulcher and Davis\cite{Fulcher} using a pendulum made with piano
wire (measuring two successive swings) and by Curtis\cite{Curtis}
who determined the period as the average of ten successive periods
for each initial amplitude. Both papers are good examples of
accurate period measurements made with an ordinary stopwatch. The
measurement of the time interval for $n$ successive periods is a
good strategy for oscillations in the small angle regime, where the
amplitude does not change significantly from a swing to the next,
but not for large-angle oscillations, because the period decreases
considerably due to air friction. This behavior is confirmed in
Fig.~4, where the period $T$ in units of $T_{0}$ is plotted as a
function of $\theta_{0}$. In Fig.~4 the curves for each
approximation formula discussed in Sec.~\ref{sec:compare} are
plotted. Experimental data taken from Refs.~\onlinecite{Fulcher} and
\onlinecite{Curtis} and the measurements taken by us in a more
sophisticated experiment\cite{Arun,pprint} are also shown. The
experimental data for amplitudes greater than $35^{\circ}$ clearly
reveal a systematic overestimation for the period due to air
damping.

In our experiment both the time-keeping and position detection
were done automatically to reduce the instrumental error to milliseconds,
which is much less than the error in time keeping when a common stopwatch
is used (of the order of 0.2\,s, the average human reaction time).
We measured the pendulum period by measuring the time interval between two
successive passages of the pendulum over the lower point of its circular
path, which corresponds to $T/2$. The measurement was based on the
variation of the electrical resistance of a light-dependent resistor
during the passing of the pendulum's bob through the path of the light from
a laser.\cite{Arun} Electronic circuitry
is needed for converting the analogue signal generated in the
light-dependent resistor when the pendulum's bob cuts the light's path
to TTL compatible digital voltage, so that the microprocessor can
understand the change in current. The details of the
design/operation of the circuitry and the microprocessor program required
for measuring the time interval between successive interruptions in the
light-dependent resistor are in
Ref.~\onlinecite{pprint}.

We devoted much attention to the reduction of the air resistance on
the motion of the pendulum bob by choosing suitable materials and
parameters for the pendulum. We used lead as the bob material due to
its high density in comparison to other inexpensive metals, which
gives a small size and large weight for the bob. We found that a
cylinder is preferable to a sphere because it allows for a better
localization of the center of mass, which is needed for measuring
$L$ accurately. The cylindrical shape also yields a reduction of air
resistance by reducing the scattering cross-section, that is, by
choosing a diameter much smaller than the height of the cylinder.
These considerations led us to fabricate a body with a mass of
0.400\,kg. For this massive bob we verified that cords made of
nylon, a commonly used material, are inadequate because they stretch
considerably for large angle oscillations and cause undesirable
vibrations. The more convenient material taking into account low
elasticity, lightness (see Ref.~\onlinecite{Armstrong} for the
importance of this factor), price, and availability, seems to be
cotton. We used a common sewing thread as the pendulum cord. We also
investigated what cord length would give the best experimental
results for large angles.\cite{length} After comparing many lengths
for an amplitude of $60^{\circ}$, we choose a length of 1.50\,m so
that the bob speed would be small (because air friction increases
with speed, the longer the string length, the less the effect of air
friction on the period) This length has a period of $\approx
2.5$\,s, which is sufficiently small for doing several repetitions
of the period measurement for each amplitude during a typical
one-hour class. These considerations led us to much more accurate
experimental data for the pendulum period for amplitudes less than
or equal to $90^{\circ}$ as shown in Fig.~4. It is seen that our
experimental data (black diamonds) are closer to the exact period
expected in the absence of air resistance (the solid line) than the
data in Refs.~\onlinecite{Fulcher} (crosses) and \onlinecite{Curtis}
(circles). The logarithmic formula in Eq.~\eqref{eq:log} is also in
better agreement with the experimental data.

\begin{acknowledgments}
F.M.S.L.\ thanks Prof.\ Get\'{u}lio T.\ Brasil (UniCEUB, Bras\'{\i}lia-DF)
for assistance on the experimental setup.
\end{acknowledgments}

\newpage

\section*{Figure Captions}

\begin{figure}[h!]
\begin{center}
\scalebox{0.60}{\includegraphics{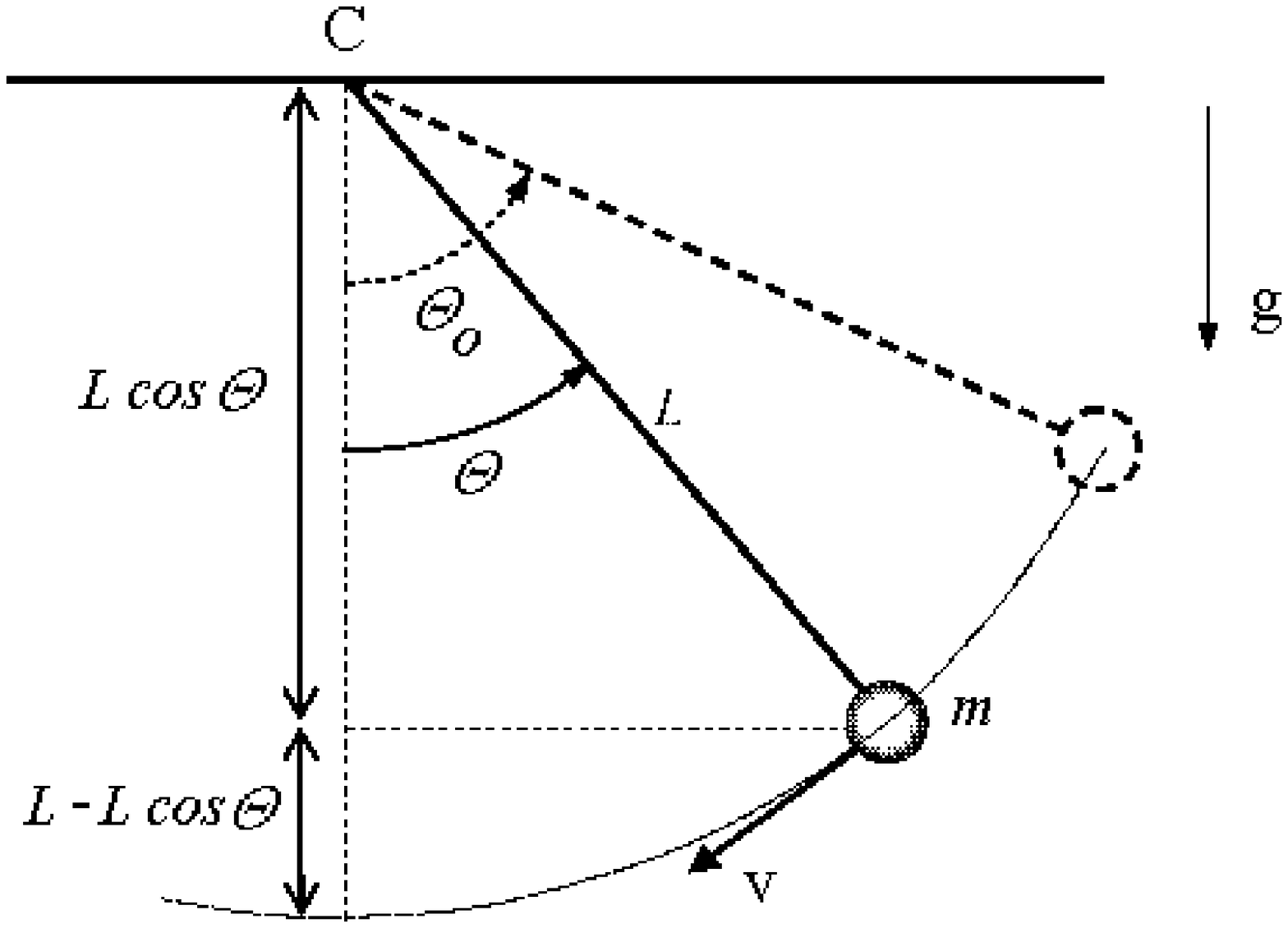}}
\caption{\label{fig:pendu} The pendulum bob is released at rest from
a position that forms an angle $\theta_{0}$ with the vertical and
passes at an arbitrary angle $\theta$ ($<\theta_{0}$) with a
velocity $L\,d\theta/dt$. Its height depends on $\theta$ according
to $L(1-\cos\theta)$.}
\end{center}
\end{figure}

\begin{figure}[h!]
\begin{center}
\scalebox{1.25}{\includegraphics{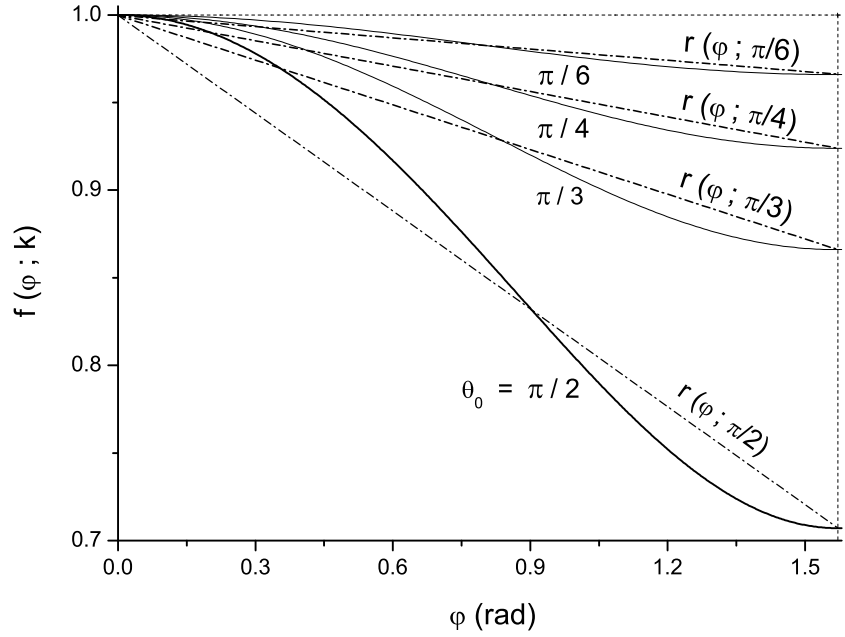}}
\caption{\label{fig:fphik} Behavior of the function
$f(\varphi,k)=\sqrt{1-k^2\sin^2\varphi}$ for $0 \leq \varphi
\leq\pi/2$ rad and for some values of $\theta_{0}$
($k=\sin(\theta_{0}/2)$). The horizontal and vertical dashed lines
are for $f(\varphi,k)=1$ and $\varphi=\pi/2$ rad, respectively. The
dash-dotted lines are the linear interpolation in Eq.~\eqref{r} for
$\theta_{0} = \pi/6$, $\pi/4$, $ \pi/3$, and $\pi/2$ rad.}
\end{center}
\end{figure}

\begin{figure}[h!]
\begin{center}
\scalebox{1.25}{\includegraphics{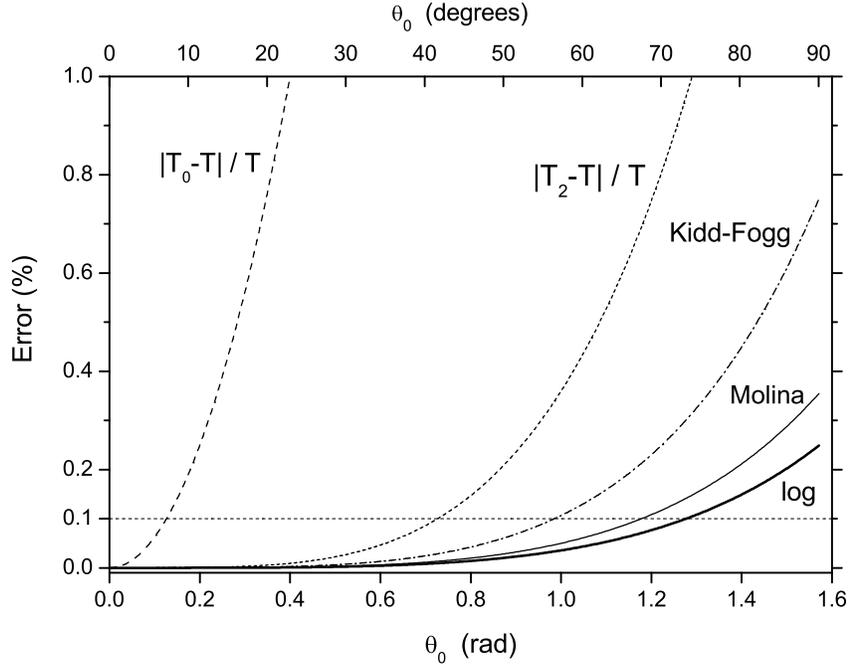}}
\caption{Comparison of the relative errors for the various approximations
discussed in the text for the period. All curves increase
monotonically with $\theta_{0}$. The horizontal dashed line marks the 0.1\%
level. The small angle approximation ($T \approx T_{0}$) yields an error
that is greater than 0.1\% for $\theta_{0}>7^\circ$ and reaches 15.3\% for
$\theta_{0}>90^\circ$. The thick solid line is for Eq.~\eqref{eq:log}. Note
 that it remains below all other curves for
$0^\circ \leq \theta_{0} \leq 90^\circ$.}
 \label{fig:relerror}
\end{center}
\end{figure}

\begin{figure}[h!]
\begin{center}
\scalebox{1.25}{\includegraphics{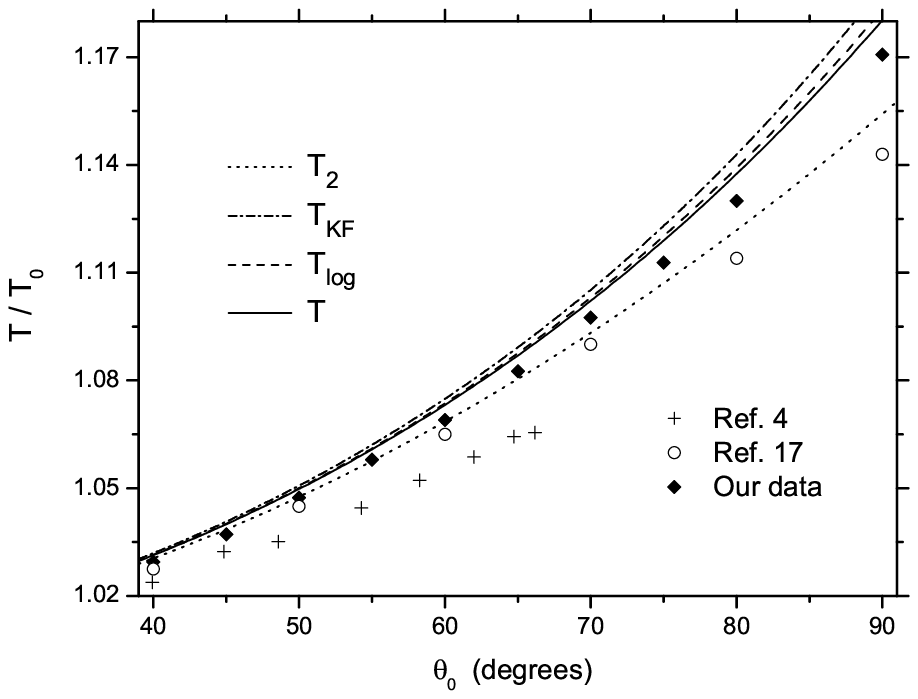}}
\caption{\label{fig:Texperim} Comparison of the ratio $T/T_{0}$ for
the approximation formulas discussed in the text and experimental
data. The dotted curve is for the Bernoulli formula, Eq.~(11). The
dash-dotted curve is for the Kidd-Fogg formula, Eq.~(12). The dashed
line is for our logarithmic formula, Eq.~\eqref{eq:log}. The solid
line is the curve for the exact period, found by numerical
integration of $K(k)$. The experimental data were taken from
Ref.~\onlinecite{Fulcher} ($+$) and Ref.~\onlinecite{Curtis}
($\circ$), and the black diamonds are our experimental data.}
\end{center}
\end{figure}


\begin{thebibliography}{24}
\bibitem{obs1} This error depends on $\theta_{0}$ implicitly (through $k$) and
 its absolute value increases rapidly with it. For instance, $T_{0}$
underestimates the exact period with an error of 15.3\% for an
amplitude of ${\! \pi/2}$ rad.
\bibitem{obs2} Of course, the cases with $\theta_{0} > {\! \pi/2}$ rad are of
less interest because most simple pendulum experiments in
introductory physics labs are done with flexible cords instead of
rigid rods, which prevents the pendulum bob from following a
circular path soon after it is released. However, our approximate
formula is more accurate than other ones even for $\theta_{0} >
\pi/2$ rad.
\bibitem{obs3} Note that the error with respect to the exact period $T$, for each amplitude, is,
 naturally, the quantity to be analyzed here instead of the error with respect to $T_{0}$.
\bibitem{length} The lengths were measured after tying up the thread firmly to a hook in the
 ceiling lab, at one end, and to a small ring at the top of the lead cylinder, at the other end.
\end{thebibliography}
\end{document}